\documentclass[twocolumn]{aastex62}

\graphicspath{{./}{figures/}}

\accepted{for publication in ApJ, 2020 November 27}

\shorttitle{The Fundamental Plane of MQGs at $z\sim2$}
\shortauthors{Stockmann et al}

\begin{document}

\title{The Fundamental Plane of Massive Quiescent Galaxies at $\MakeLowercase{z}\sim2$}

\author[0000-0001-5983-6273]{Mikkel Stockmann}
\affil{Cosmic Dawn Center (DAWN), Denmark}
\affil{Niels Bohr Institute, University of Copenhagen, Jagtvej 128, 2100 Copenhagen Ø, Denmark}
\affil{DARK, Niels Bohr Institute, University of Copenhagen, Jagtvej 128, DK-2100 Copenhagen, Denmark}
\email{mikkelstockmann@gmail.com}

\author[0000-0003-3002-1446]{Inger J{\o}rgensen}
\affil{Gemini Observatory, 670 N.\ A`ohoku Pl., Hilo, HI 96720, USA}

\author[0000-0003-3631-7176]{Sune Toft}
\affil{Cosmic Dawn Center (DAWN), Denmark}
\affil{Niels Bohr Institute, University of Copenhagen, Jagtvej 128, 2100 Copenhagen Ø, Denmark}

\author[0000-0003-1949-7638]{Christopher J. Conselice}
\affil{Centre for Astronomy and Particle Theory, School of Physics $\&$ Astronomy, University of Nottingham, Nottingham, NG7 2RD, UK}
\affiliation{Jodrell Bank Centre for Astrophysics, University of Manchester, Oxford Road, Manchester UK}

\author[0000-0002-9382-9832]{Andreas Faisst}
\affil{IPAC, California Institute of Technology, 1200 East California Boulevard, Pasadena, CA 91125, USA}

\author[0000-0001-8702-7019]{Berta Margalef-Bentabol}
\affil{LERMA, Observatoire de Paris, PSL Research University, CNRS, Sorbonne Universit\'es, UPMC Univ. Paris 06}

\author[0000-0002-9656-1800]{Anna Gallazzi}
\affil{INAF-Osservatorio Astrofisico di Arcetri, Largo Enrico Fermi 5, I-50125 Firenze, Italy}

\author[0000-0003-1734-8356]{Stefano Zibetti}
\affil{INAF-Osservatorio Astrofisico di Arcetri, Largo Enrico Fermi 5, I-50125 Firenze, Italy}

\author[0000-0003-2680-005X]{Gabriel B. Brammer}
\affil{Cosmic Dawn Center (DAWN)}
\affil{Niels Bohr Institute, University of Copenhagen, Jagtvej 128, 2100 Copenhagen Ø, Denmark}

\author[0000-0002-4085-9165]{Carlos G\'omez-Guijarro}
\affil{AIM, CEA, CNRS, Universit\'e Paris-Saclay, Universit\'e Paris Diderot, Sorbonne Paris Cit\'e, F-91191 Gif-sur-Yvette, France}

\author[0000-0002-3301-3321]{Michaela Hirschmann}
\affil{DARK, Niels Bohr Institute, University of Copenhagen, Jagtvej 128, DK-2100 Copenhagen, Denmark}

\author[0000-0003-3021-8564]{Claudia D. Lagos}
\affil{Cosmic Dawn Center (DAWN)}
\affil{International Centre for Radio Astronomy Research (ICRAR), M468, University of Western Australia, 35 Stirling Hwy, Crawley, WA 6009, Australia}
\affil{ARC Centre of Excellence for All Sky Astrophysics in 3 Dimensions (ASTRO 3D)}

\author[0000-0001-6477-4011]{Francesco M. Valentino}
\affil{Cosmic Dawn Center (DAWN)}
\affil{Niels Bohr Institute, University of Copenhagen, Jagtvej 128, 2100 Copenhagen Ø, Denmark}
\affil{DARK, Niels Bohr Institute, University of Copenhagen, Jagtvej 128, DK-2100 Copenhagen, Denmark}

\author[0000-0002-9842-6354]{Johannes Zabl}
\affil{Univ Lyon, Univ Lyon1, Ens de Lyon, CNRS, Centre de Recherche Astrophysique de Lyon UMR5574, F-69230 Saint-Genis-Laval, France}

\begin{abstract}

We examine the Fundamental Plane (FP) and mass-to-light ratio ($M/L$) scaling relations using the largest sample of massive quiescent galaxies at $1.5<z<2.5$ to date. The FP ($r_{e}, \sigma_{e}, I_{e}$) is established using $19$ \textit{UVJ} quiescent galaxies from COSMOS with \textit{Hubble Space Telescope (HST)} $H_{F160W}$ rest-frame optical sizes and X-shooter absorption line measured stellar velocity dispersions. For a very massive, ${\rm{log}}(M_{\ast}/M_{\odot})>11.26$, subset of 8 quiescent galaxies at $z>2$, from Stockmann et al. (2020), we show that they cannot passively evolve to the local Coma cluster relation alone and must undergo significant structural evolution to mimic the sizes of local massive galaxies. The evolution of the FP and $M/L$ scaling relations, from $z=2$ to present-day, for this subset are consistent with passive aging of the stellar population and minor merger structural evolution into the most massive galaxies in the Coma cluster and other massive elliptical galaxies from the MASSIVE Survey. Modeling the luminosity evolution from minor merger added stellar populations favors a history of merging with ``dry'' quiescent galaxies.
\end{abstract}





\section{Introduction}
The most massive local elliptical galaxies, believed to be one of the most mature stages of galaxy evolution, have been shown to form the majority of their stars rapidly, in the densest environments at $z>2-5$ \citep[e.g.][]{Blakeslee+03,Thomas+05,Greene+15}. Understanding the formation and evolution of these systems is a complex task. One way to address this is to study their progenitors in the early universe and to see how their properties differentiate from their $z=0$ counterparts, from which evolution can be inferred.

A population of massive, ${{\rm{log}}}(M_{\ast}/M_{\odot})>11$, quiescent galaxies which are the possible progenitor candidates of modern ellipticals have been located at $z>2$ \citep{Daddi+04,Kriek+09,Toft+12,Belli+17}, which allows us to observe the evolution of the most massive systems in the Universe. It is clear that these early massive galaxies have extremely compact sizes \citep{Daddi+05,Trujillo+06,Trujillo+07,Buitrago+08,vanDokkum+08,Conselice+11,Szomoru+12}, which are 3-5 times smaller than the present-day most massive elliptical galaxies at the same mass, and are also younger with more recent star formation (e.g., \citealt{Stockmann+20}).

From these observations rapid size evolution has been inferred for field early-type galaxies across time \citep{Newman+12,vanderWel+14,Faisst+17,Mowla+18,Morishita+18}. Simulations have shown dry mergers to be an efficient process for making galaxies larger \citep[e.g.][]{Hopkins+09,Naab+09,Bezanson+09,Hilz+12,Hilz+13,Remus+17,Lagos+18}. \cite{Toft+14} proposed an evolutionary sequence of massive galaxies where the most massive elliptical galaxies, from the present-day Universe, were formed in violent star-bursts. These later quench possibly via AGN to become the compact quiescent galaxies at $z>2$ suggested to undergo rapid size evolution and become the massive elliptical galaxies in the local universe \citep[see also][]{Cimatti+08a,Simpson+14,Gomez-Guijarro+18,Habouzit+19}.

Scaling relations between different properties of galaxies, and how these evolve through time can give us significant information about how galaxies are assembled over cosmic time.   One way to do this is by studying the evolution of scaling relationships between various quantities.  For example, massive local elliptical galaxies in the nearby Universe are found to follow an empirical relation known as the Fundamental Plane between surface brightness, internal velocity, and size \citep[FP,][]{Djorgovski&Davis+87,Dressler+87}.

The zero point of the edge-on FP has been observed to evolve with redshift, complementary to the $M/L$ ratio \citep{Faber+87}, which has made this a preferred tool in studying the structural and luminosity evolution of early-type galaxies across time \citep[e.g.][]{Bender+92,Jorgensen+96,Jorgensen+99,Treu+05,vanderWel+05,Cappellari_2006,Jorgensen+06,vanderMarel_vanDokkum+07,Saglia+10,Jorgensen_Chiboucas2013}.   At $z<1$, the FP zero point offset has been interpreted as the result of purely passive (without structural) evolution of the stellar population \citep[e.g.][]{Jorgensen+06,Jorgensen_Chiboucas2013}. However, this is likely not the case at $z>2$ where the red and quiescent galaxies are compact and must undergo significant size evolution to evolve into the sizes of the present-day galaxies.

Spectroscopic observations which are required to measure stellar velocity dispersions at $z>2$ are however time-expensive and only the rarest, brightest and most massive systems have been studied at this distance \citep{vandeSande2013,Belli+17} using large cosmological fields like CANDELS \citep{Grogin+11,Koekemoer+11} and COSMOS \citep{Scoville+07}.

We present in this paper the FP study at $z>2$ using a sample of massive field quiescent galaxies (MQGs) introduced in \cite{Stockmann+20} (hereafter S20). S20 find a shallow stellar velocity dispersion evolution and significant size growth between $z=2$ and $0$. In this paper, we explore whether this size growth, alongside the passive evolution of the stellar population, can account for the observed evolution of massive galaxies in the scaling relations from $z=2$ to the present-day.

In Section \ref{sec:data}, the $z>2$ MQGs sample from S20 together with a complementary quiescent galaxy sample at a similar redshift is presented alongside two local samples from the Coma cluster and the MASSIVE Survey. We present the $M/L$ and FP scaling relations in Section \ref{sec:ML_M_sigma} and \ref{sec:FP}, respectively. The predicted evolution of the size, stellar velocity dispersion, passive aging, and luminosity increase due to minor merger driven growth are presented in Section \ref{sec:SR_evol}. Finally, our results are interpreted and discussed in Section \ref{se:discussion}, following a summary of the main conclusions in Section \ref{sec:summary}.\\

Throughout the text, magnitudes are quoted in the AB system \citep{Oke_n_Gunn+83,Fukugita+96} and the following cosmological parameters, $\Omega_{m}=0.3$, $\Omega_{\Lambda}=0.7$, with $H_{0}=70\ \rm{km}\ \rm{s}^{-1}\ \rm{Mpc}^{-1}$ are used. All stellar masses are presented using the \cite{Chabrier+03} Initial Mass Function (IMF).

\begin{table*}[th!]
\caption{Sample Summary}
\label{tab:data}
\begin{tabular}{llllllllll}
\hline
Target ID & RA & Dec &  $z_{\rm{spec}}$  & $\mathrm{log}\rm{Age/yr}$ & ${\rm{log}}M_{\ast}/M_{\odot}$ & ${\rm{log}}M_{\rm{dyn}}/M_{\odot}$ & $r_{e, circ}$ & $\sigma_{e}$ & ${\rm{log}}\langle I\rangle_{e,\mathrm{B}}$  \\ \hline
UV-105842              & 150.26265   & 2.0177791   & 2.0195    & $9.19^{+0.26}_{-0.33}$    & $11.68^{+0.16}_{-0.17}$   & $11.47\pm0.19$   & $2.91\pm0.29$    & $263\ \pm57$        & 4.40    \\
UV-171687              & 149.88702   & 2.3506956   & 2.1020    & $9.13^{+0.28}_{-0.32}$    & $11.51^{+0.18}_{-0.19}$   & $11.31\pm0.24$   & $4.49\pm0.45$    & $182\ \pm50$        & 3.99    \\
UV-90676               & 150.48750   & 2.2700379   & 2.4781    & $9.09^{+0.29}_{-0.29}$    & $11.78^{+0.17}_{-0.18}$   & $11.78\pm0.21$   & $4.08\pm0.41$    & $347\ \pm82$        & 4.46    \\ 
UV-155853              & 149.55630   & 2.1672480   & 1.9816    & $9.23^{+0.24}_{-0.33}$    & $11.62^{+0.18}_{-0.17}$   & $11.57\pm0.11$   & $4.20\pm0.42$    & $247\ \pm30$        & 3.96    \\
UV-230929              & 150.20842   & 2.7721019   & 2.1679    & $9.10^{+0.28}_{-0.28}$    & $11.48^{+0.16}_{-0.16}$   & $11.16\pm0.07$   & $1.48\pm0.15$    & $252\ \pm21$        & 5.11    \\ 
CP-1243752             & 150.07394   & 2.2979755   & 2.0903    & $9.23^{+0.24}_{-0.32}$    & $11.79^{+0.17}_{-0.17}$   & $11.61\pm0.13$   & $2.54\pm0.25$    & $350\ \pm53$        & 4.59    \\ 
CP-540713              & 150.32512   & 1.8185385   & 2.0409    & $9.16^{+0.27}_{-0.32}$    & $11.26^{+0.22}_{-0.23}$   & $11.53\pm0.24$   & $1.46\pm0.15$    & $353\ \pm97$        & 4.66    \\
UDS-19627$^\mathrm{a}$ & 34.57125    & -5.3607778  & 2.0389    & $9.08^{+0.11}_{-0.10}$    & $11.37^{+0.13}_{-0.10}$   & $11.33\pm0.14$   & $1.43\pm0.14$    & $318\ \pm53$        & 5.20    \\
\hline
\end{tabular}

\tablecomments{Column 1: Target ID from S20; Column 2: Right Ascension in degrees (J2000); Column 3: Declination in degrees  (J2000); Column 4: Spectroscopic redshift; Column 5: Mass-weighted Age; Column 6: Stellar mass; Column 7: Circularised dynamical mass calculated using $\beta{(n)}$ from \cite{Cappellari+06}; Column 8: Circularised effective radius in kpc; Column 9: Stellar velocity dispersion in ${\rm{km}}\ {\rm{s}}^{-1}$; Column 10: Average Surface brightness within $r_{e, circ}$ in ${\rm{L}}_{{\rm{B}},\odot{}}{\rm{pc}}^{-2}$ (see Appendix \ref{app:Ie}).}

\tablenotetext{a}{The spectroscopic redshift, age, stellar mass, and stellar velocity dispersion are from \cite{Toft+12} and the \textit{HST}/WFC3 $H_{\rm{F160W}}$ size are from S20.} 

\end{table*}


\section{Data} \label{sec:data}

\subsection{A sample of massive quiescent galaxies at $z>2$} \label{sec:S20}

In S20, we presented a sample of MQGs at $z>2$ studied with the X-shooter spectrograph \citep{D'odorico+06,Vernet_2011_Xshooter} at the \textit{VLT} and \textit{Hubble Space Telescope} (\textit{HST}) $H_{F160W}$ that are crucial to obtain both rest-frame optical stellar velocity dispersions and effective sizes. The sample is selected from the 2 square degree COSMOS field, using multi-waveband photometric fits \citep{Muzzin_COSMOS_Uvista}. In summary, the sample is selected to be \textit{K}-band bright and massive (${\rm{log}}(M_{\ast}/M_{\odot})>11$) \textit{UVJ} quiescent galaxies at $z>2$. The adopted stellar population parameters such as mass-weighted age and stellar mass were based on the COSMOS15 photometry \cite{Laigle+16}, the continuum emission modelling of the X-shooter spectra, and the choice of star-formation history (see details in S20 section 4.3). Here we consider 8 of the total 15 galaxies from S20 with measured stellar velocity dispersions, which is essential to study them in the scaling relations. We find no selection bias when comparing the size, age, stellar mass, and redshift of this sample to the parent sample in S20.\\

In Table \ref{tab:data}, we list the mass-weighted age, stellar mass, stellar velocity dispersions, and sizes for these 8 galaxies. The adopted stellar velocity dispersion will be referred to as the effective stellar velocity dispersion in this study due to minimal correction ($<5\%$) when following the equations based on X-shooter observations in \cite{vandeSande2013}. For further details, we refer to S20. Contrary to S20, we compute the dynamical masses in this paper by using the circularized sizes ($r_{e,circ} = \sqrt{ba}$)\footnote{Here $a$ and $b$ are the semi-major and -minor axis.} to make them consistent with the dynamical masses derived using circularized radius from the study of local cluster scaling relations in previous work such as e.g. \cite{Jorgensen_Chiboucas2013}. We have verified that the qualitative results from S20 remain when using circularized dynamical masses and sizes. Following local studies \citep[e.g.][]{Jorgensen+99}, we adopt the Bessel B-band luminosity, estimated in our case from the rest-frame fluxes obtained from the most recent COSMOS photometry \citep{Laigle+16} using the photometric redshift code EAZY \citep{Brammer+08}\footnote{\url{https://github.com/gbrammer/eazy-photoz}}. Normally, the magnitudes (luminosities) from the effective radius profile fitting is used when studying the scaling relations. We adopted another approach to, instead, extract the COSMOS magnitudes to treat the high redshift samples (S20 and the galaxies from Section \ref{sec:belli17}) consistently. We confirm that these magnitudes, when compared to the $H_{F160W}$ magnitudes from the profile fits, can be considered representative with negligible differences. The method here is as a result similar to the standard approach. The luminosity and average effective surface brightness are estimated using the method outlined in Appendix \ref{app:Ie}. Hereafter, the dynamical mass-to-light ratio in the Bessel B-band is referred to as $M/L$.
The $8$ galaxies in our sample have a mean age of $\sim1.5$ Gyr and span a similar stellar mass and size range to the full 15 galaxy parent sample from S20.


\subsection{Complementary sample of quiescent galaxies at $1.5<z<2.5$} \label{sec:belli17}
In addition to the galaxies from S20, we adopt a sample of $1.5<z<2.5$ quiescent galaxies from \cite{Belli+17}. We choose 11 out of 24 galaxies with stellar velocity dispersions, which have available COSMOS photometry in \cite{Laigle+16} to ensure consistent photometry extractions similar to the S20 sample. Out of the 11 galaxies, 7 are at $z<2$ and 4 at $z>2$. The $11$ galaxies from \cite{Belli+17} introduced here will be referred to as B17.

From \cite{Belli+17}, we adopt the effective semi-major axis, absorption line measured stellar velocity dispersions and dynamical masses (see their table 2). The effective semi-major axis are derived, similar to S20, using S\'ersic profile fits to the rest-frame optical \textit{HST}/WFC3 $H_{\rm{F160W}}$ images using \textit{GALFIT} \citep{Peng_Galfit} and converted to circularized radius. We know from S20 sample that the COSMOS magnitudes in the H-band are a good approximation of the magnitudes in the above profile fit method and thus indirectly related to the effective radii. The stellar velocity dispersions are derived from rest-frame optical Multi-Object Spectrometer For Infra-Red Exploration (MOSFIRE, \citealt{McLean_MOSFIRE_2012}) spectra using Penalized PiXel-Fitting (pPXF, \citealt{Cappellari_ppxf}) to fit \cite{BC+03} (BC03) stellar population models which are qualitative similar to S20. The dynamical masses are converted from the semi-major axis to the circularized radius using the axis ratio. As in S20, the dynamical mass is estimated using the method from \cite{Cappellari_2006}.\\

The two $z\sim2$ samples (Section \ref{sec:S20} and \ref{sec:belli17}), used here, satisfy the \textit{UVJ} quiescent galaxy selection from \cite{Muzzin_COSMOS_Uvista}. Together they allow for exploring a larger dynamical mass range, $10.5<{\rm{log}}(M_{\rm{dyn}}/M_{\odot})<11.9$, and redshift, $1.5<z<2.5$, corresponding to a cosmological time span of $\sim1.7$ Gyr considered short compared to their consequent 8-10 Gyrs of evolution to the present-day. 


\subsection{The MASSIVE Survey} \label{sec:MASSIVEnx}
The original volume-limited MASSIVE Survey sample is selected as the most massive and \textit{K}-band brightest early-type galaxies within $108$ Mpc of the northern hemisphere \citep{Ma+14}. Here, we use the 25 most massive, ${\rm{log}}(M_{\ast}/M_{\odot})>11.7$, MASSIVE galaxies (hereafter MASSIVE(n) sample), selected at fixed cumulative number density (CND) matching our massive $z\sim2$ sample as described in S20. The cumulative number density of the $z\sim2$ sample is estimated from the massive, ${\rm{log}}(M_{\ast}/M_{\odot})>11.2$, \textit{UVJ} quiescent galaxies at $1.9<z<2.5$ in the \cite{Muzzin_COSMOS_Uvista} catalog. The results in S20 are shown to be robust against the choice of CND method (fixed and probabilistic, \citealt{Wellons&Torrey+17}), as well as the mass-rank scatter. Utilizing this CND approach is an attempt to minimize progenitor bias by predicting the local progenitors of the high-z galaxies. A thorough discussion of the assumptions and uncertainties is covered in S20 Section 5.1.

The magnitudes for 17/25 galaxies are obtained from the SDSS DR14 catalog photometry \citep{Blanton+17_SDSS_IV} by cross-matching the MASSIVE(n) sample using the SDSS SkyServer\footnote{\url{http://skyserver.sdss.org/dr14/en/tools/crossid/crossid.aspx}}. The {\textit{de Vaucouleur}} photometry (``deVMag'') in the \textit{u, g, r, i, z} bands is extracted and converted to rest-frame Bessel B-band magnitude using the EAZY code, in the same way as in the high redshift samples in this study. The ``cmodelMag'' estimate, also used for the Coma sample (Section \ref{sec:coma_data}), was compared with the {\textit{de Vaucouleur}} fit magnitudes to establish that the latter is a good representation, which was also confirmed by the ``fracDeV'' parameter. The luminosity and effective surface brightness are calculated from the apparent magnitude using the methods covered in Appendix \ref{app:Ie}. Instead of estimating the luminosity distance from the redshift, these galaxies are close enough that peculiar velocities have a significant impact on their distance measurement. We, therefore, use the distance measurement from \cite{Ma+14}, who correct for this effect. The $17$ galaxy subsample used here is referred to as MASSIVE(n$_{17}$).\\

The {\textit{de Vaucouleur}} effective radii (``deVRad'') corresponding to extracted photometry from the SDSS SkyServer is used. After confirming that the radii are consistent among the different bands, we chose to use the $g$-band effective radii as the wavelength coverage is comparable to the Bessel B-band. The radii are circularized using the axis ratio (``deVAB"). The average luminosity weighted dispersion within the effective radius is adopted \citep{Veale+18}. The dynamical masses are estimated using the method in S20 \citep[with the prescription from][]{Cappellari+06} using $n=4$ and the circularized effective sizes.\\

The stellar mass, size and stellar velocity dispersion between the MASSIVE(n) and MASSIVE(n$_{17}$) samples are compared in Appendix \ref{app:massiven}. Here we find that the MASSIVE(n$_{17}$) sample is uniformly sampled from the initial distribution of stellar mass, size and stellar velocity dispersion and is $68\ \%$ complete. As a result, our MASSIVE(n$_{17}$) selection with available photometry is representative of the parent CND matched sample and can be considered a suitable minimal progenitor biased reference sample.

\begin{figure*}[t!]
    \centering
        \includegraphics[width=18cm]{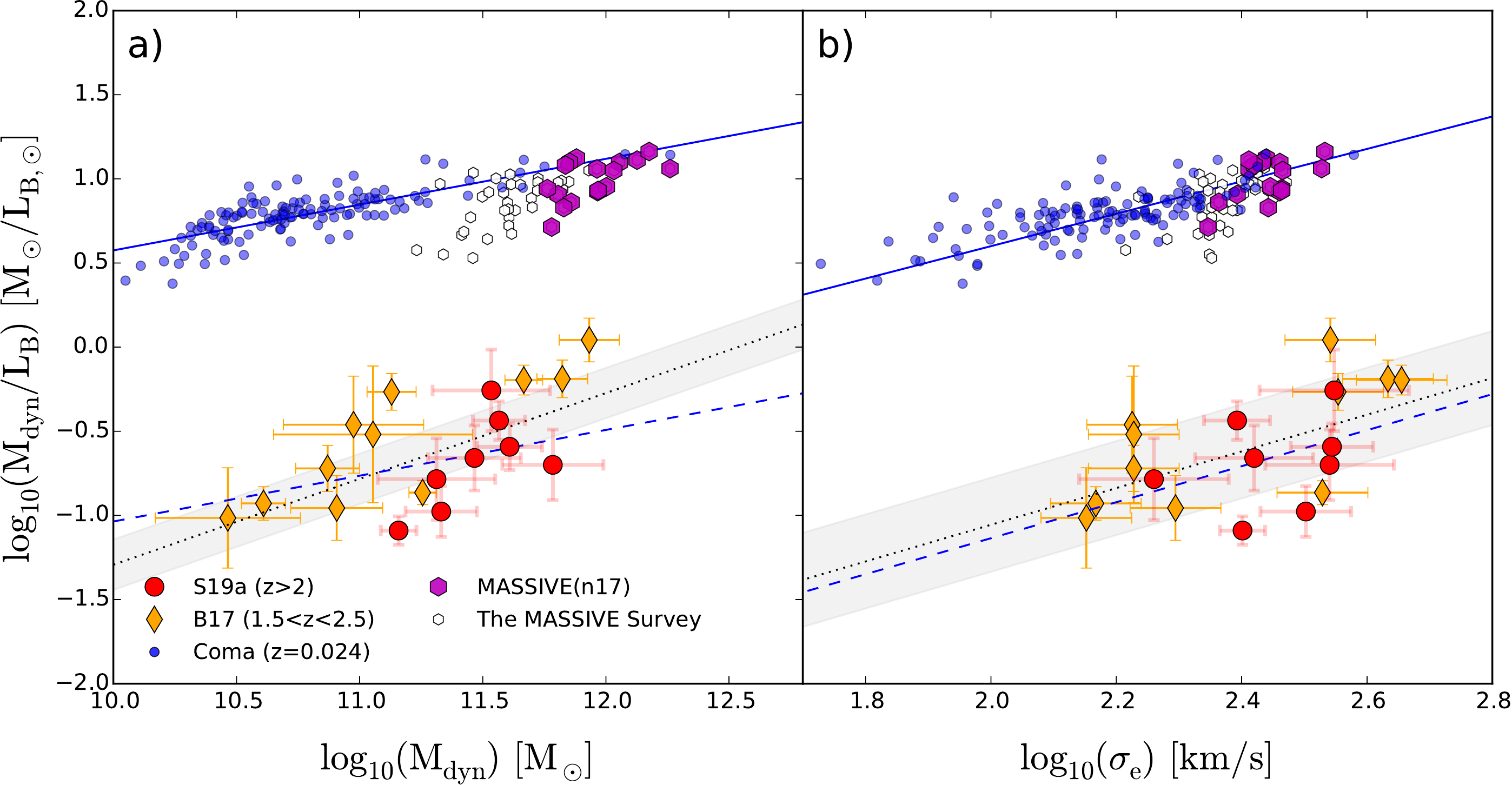}
        \caption{$M/L$ ratio as a function of dynamical mass (a) and stellar velocity dispersion (b) are shown for the S20 MQGs $z>2$ (red symbols) and $11$ COSMOS quiescent galaxies at $1.5<z<2.5$ from B17 (orange symbols). The Coma galaxies ($z=0.0231$, blue symbols) are shown together with the best-fit relation (blue line) when assuming \citealt{Jorgensen_Chiboucas2013} FP parameters. The MASSIVE Survey galaxies with available rest-frame B-band photometry (all white and purple hexagons) and the CND-matched MASSIVE(n$_{17}$) sample (purple hexagons only) are shown. The best-fit (black dotted line) and rms (gray shading) of the combined high redshift samples (S20 and B17) are shown together with the Coma best-fit relation offset to the median $M/L$ at $z\sim2$ (dashed blue).}
        \label{fig:ML_M_sigma}
\end{figure*}


\subsection{Coma cluster} \label{sec:coma_data}

As a local reference cluster we use Coma/Abell 1656 complete to $g^{\prime}_{rf}\leq16.1$ magnitudes, as used in \cite{Jorgensen+19}. The scaling relations of the Coma cluster are well studied and we include them as a reference to other high-redshift comparisons. We adopt the $123$ galaxies with average effective surface brightness, stellar velocity dispersion, and circularised sizes. The stellar velocity dispersion measurements (from \citealt{Jorgensen+18}) are derived from high S/N ($\sim60\ \mathrm{\AA{}}^{-1}$) spectra using the recipe presented in \cite{Jorgensen+17}. The surface brightness is calculated using rest-frame Bessel B magnitudes based on the SDSS \textit{cmodelmag} magnitude (see Appendix A in \citealt{Jorgensen+19}). The circularized effective radii are constructed from SDSS parameters to be a pseudo-S\'ersic effective radii with the purpose of matching the ``cmodelMag'' magnitudes (see method in \citealt{Jorgensen+19}). Both data sets are calibrated to the \textit{Legacy data} \citep{Jorgensen+99,Jorgensen_Chiboucas2013} to provide a trustworthy low redshift reference cluster. For further details on the physical properties of the Coma cluster, we refer to section 2 in \cite{Jorgensen+18}.



\section{Results} \label{sec:results}
In this section, we present the $M_{\rm{dyn}}/L_{B}$ and Fundamental Plane scaling relations as a tool to study the evolution of MQG from $z=2$ to $0$. We do so by fitting these relations at $z\sim2$ and comparing them to two local samples presented in Section \ref{sec:data}.


\subsection{Dynamical Mass-to-light ratio, $M_{\rm{dyn}}/L_{B}$} \label{sec:ML_M_sigma}

The $M_{\rm{dyn}}/L_{B}$ reveals information about how the dynamical structure of galaxies' compares to the luminosity of their stellar populations. This ratio has been found to increase in massive quiescent galaxies, believed to be driven by the non-star forming and passive evolution of their stellar population. Here, we study the $M_{\rm{dyn}}/L_{B}$ relations for massive quiescent galaxies and their evolution in the past 10 billion years. We show these relations in Figure \ref{fig:ML_M_sigma}, both as a function of dynamical mass (\ref{fig:ML_M_sigma}a) and stellar velocity dispersion (\ref{fig:ML_M_sigma}b), for a sample of $z\sim2$ MQGs (S20 + B17) alongside two local reference samples (Coma, MASSIVE(n$_{17}$)).

Compared to the local Coma and MASSIVE(n$_{17}$) galaxies, the high redshift samples (S20 and B17) have lower M/L as expected for brighter younger systems. The majority of the galaxies from S20 have, already at $z=2$, dynamical masses similar to the $\sim10\%$ most massive galaxies in the Coma cluster (see Figure \ref{fig:ML_M_sigma}a). The MASSIVE(n$_{17}$) galaxies have dynamical masses similar to the $2\ \%$ most dynamical massive Coma galaxies and show that they are among the most massive galaxies in the local Universe.
The distribution of the stellar velocity dispersions of S20 and B17 is consistent with the high-end measurements in Coma (see Figure \ref{fig:ML_M_sigma}b). When comparing the stellar velocity dispersions between the S20 and the MASSIVE(n$_{17}$) sample we find that a shallow evolution is expected between $z=2$ and the present-day. The MASSIVE(n$_{17}$) sample have high stellar velocity dispersions similar to the high end of the Coma cluster measurements.

The combined samples of S20 and B17, in Figure \ref{fig:ML_M_sigma}a and b, are fit by minimizing the least-squares in the y-direction and the uncertainty on the slopes is estimated using a bootstrap procedure \citep[see also][]{Jorgensen+96, Jorgensen_Chiboucas2013}. A relation for the combined high-redshift (S20 and B17) sample is established, while this was not possible using the narrow dynamical mass range of the S20 sample alone. The fits are shown in Figure \ref{fig:ML_M_sigma}, together with the associated root mean square (rms) from the regression, and listed in Table \ref{tab:fit}. The $M/L$ with dynamical mass relation appears slightly steeper, although by less than $2\sigma$. For both $M/L$ vs $M_{\rm{dyn}}$ and $M/L$ vs $\sigma_{e}$, we find best-fit slopes, for the combined high redshift sample, to be consistent with the local Coma relation (see relation 4 and 6 in Table \ref{tab:fit}).


\begin{figure}[t]
    \centering
        \includegraphics[width=8.5cm]{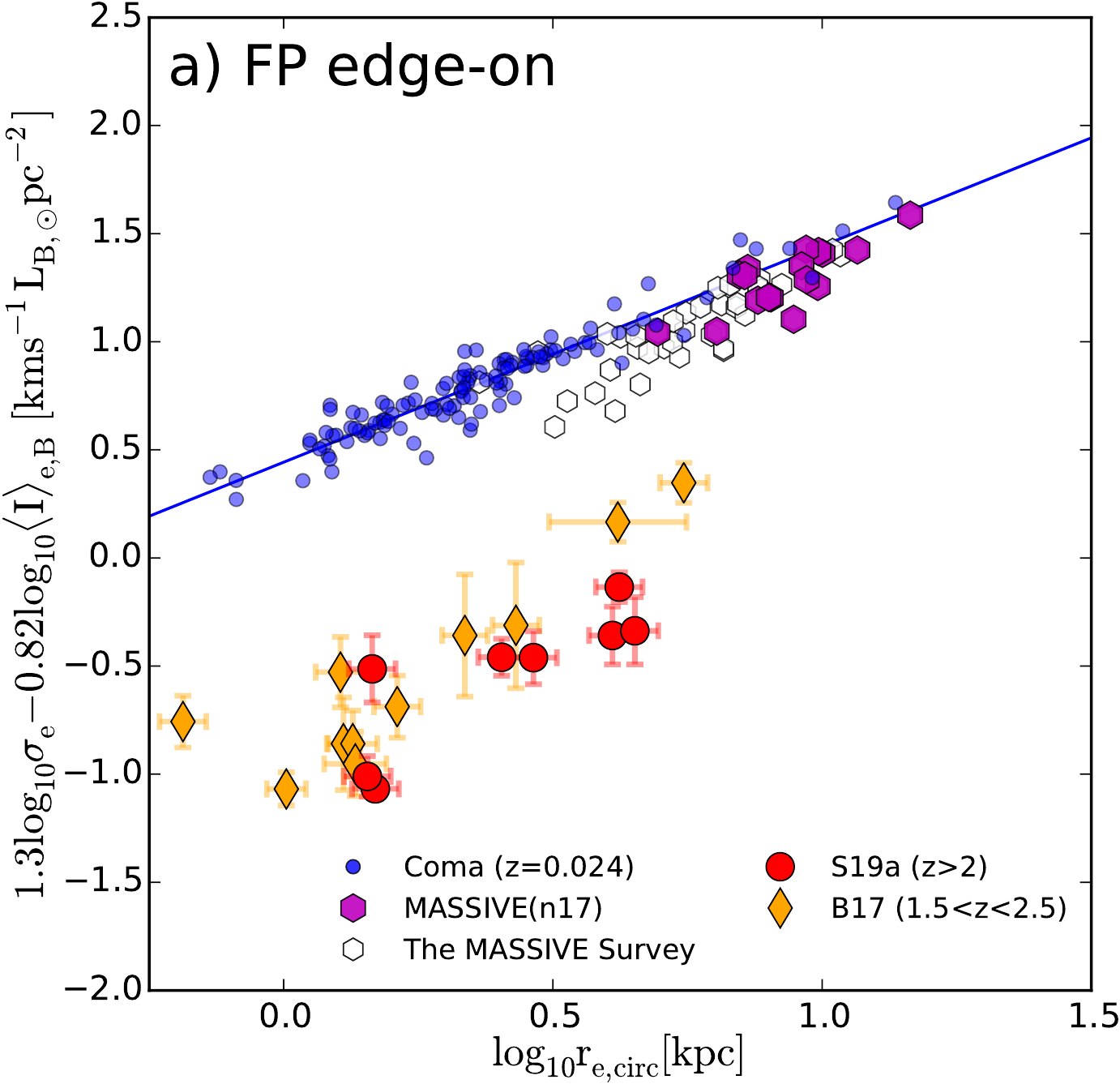}
        \includegraphics[width=8.5cm]{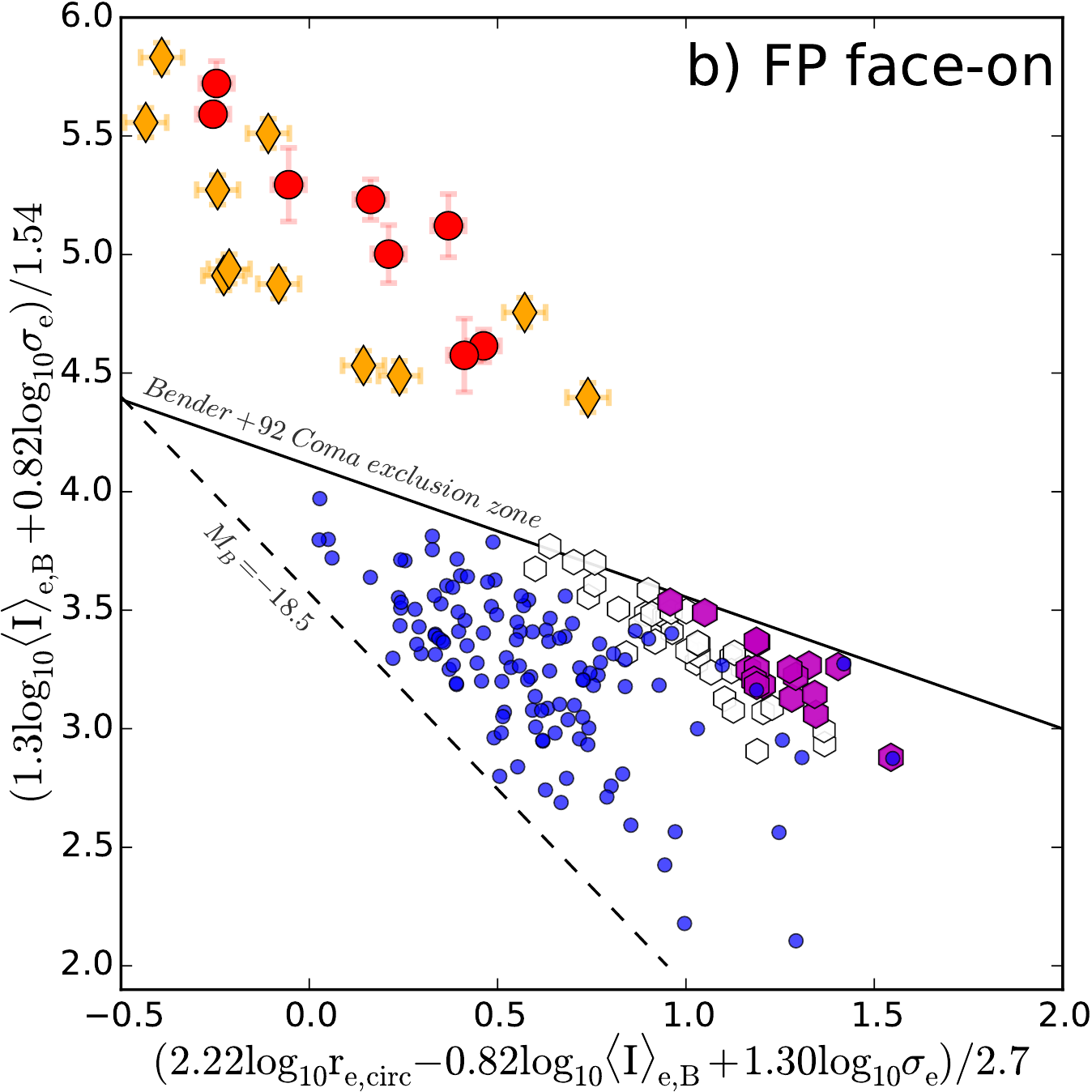}
        \caption{The FP projected edge-on (a) and face-on (b) with the symbols as in Figure \ref{fig:ML_M_sigma}. The exclusion zone for local spheroidal galaxies from \cite{Bender_1992} and the Coma luminosity limit from \cite{Jorgensen+06} are shown in the face-on plane. A $\rm{FP}_{z\sim2}$ is established when combining the COSMOS quiescent galaxies from S20 and B17.}
        \label{fig:FP_eo_fo}
\end{figure}

\begin{table*}[]
\caption{Fundamental Plane and $M/L$ scaling relations}
\label{tab:fit}
\begin{center}
\begin{tabular}{llllll}
No.  &  Sample  &  Relation  & $N_{{\rm{gal}}}$ & rms & $\sigma_{\gamma}$   \\ \hline
1) &  Coma$^{\rm{a}}$                            & $\mathrm{log}r_{e} = (1.30\pm0.08)\;\mathrm{log}\sigma - (0.82\pm0.03)\;\mathrm{log}\langle I\rangle_{e,\mathrm{B}}-0.443$  & 123   & 0.08 & 0.01 \\
2) &  Combined sample ($1.5<z<2.5$)$^{\rm{b}}$   & $\mathrm{log}r_{e} = (0.46\pm0.18)\;\mathrm{log}\sigma - (0.46\pm0.07)\;\mathrm{log}\langle I\rangle_{e,\mathrm{B}}+1.275$  & 19    & 0.15 & 0.03  \\
3) &  Coma$^{\rm{a}}$                            & $\mathrm{log}M/L = (0.27\pm0.03)\;\mathrm{log}M_{\rm{dyn}}-2.145$                                                           & 123   & 0.09 & 0.01 \\
4) &  Combined sample ($1.5<z<2.5$)$^{\rm{b}}$   & $\mathrm{log}M/L = (0.51\pm0.15)\;\mathrm{log}M_{\rm{dyn}}-6.393$                                                           & 19    & 0.26 & 0.06  \\
5) &  Coma$^{\rm{a}}$                            & $\mathrm{log}M/L = (0.96\pm0.10)\;\mathrm{log}\sigma-1.326$                                                                 & 123   & 0.10 & 0.01 \\
6) &  Combined sample ($1.5<z<2.5$)$^{\rm{b}}$   & $\mathrm{log}M/L = (1.09\pm0.41)\;\mathrm{log}\sigma-3.236$                                                                 & 19    & 0.28 & 0.06
\end{tabular}
\end{center}
\tablecomments{Column 1: Sample; Column 2: Fitting method; Column 3: Scaling relations; Column 4: Number of galaxies included in fit; Column 5: rms scatter along the y direction of the scaling relation; Column 6: The uncertainty on the zero point ($\gamma$) estimated as rms/$\sqrt{N_{{\rm{gal}}}}$ (assuming fixed coefficients).}
\tablenotetext{a}{Fits are from \cite{Jorgensen+19} Table 4.}
\tablenotetext{b}{Samples from S20 and B17 (see Section \ref{sec:data})}
\end{table*}

\subsection{The Fundamental Plane} \label{sec:FP}

The FP is spanned by the effective size, $r_{e}$, stellar velocity dispersion, $\sigma_{e}$, and average effective surface brightness, $\langle{} I\rangle{}_{e,\mathrm{B}}$. Its edge-on and face-on orientations are defined as
\begin{equation} \label{eq:FPeo}
    \mathrm{log}_{10}\; r_{e} = \alpha\; \mathrm{log}_{10}\;\sigma_{e} +\beta\; \mathrm{log}_{10}\;\langle{} I\rangle{}_{e,\mathrm{B}} + \gamma
\end{equation}
and
\begin{eqnarray} \label{eq:FPfo}
    & (2.22\;\mathrm{log}_{10}\;r_{e}+\beta\;\mathrm{log}_{10}\;\langle{} I\rangle{}_{e,\mathrm{B}}+\alpha\;\mathrm{log}_{10}\;\sigma_{e})/2.7\nonumber \\
    & = (\alpha\;\mathrm{log}_{10}\;\langle{} I\rangle{}_{e,\mathrm{B}} -\beta\;\mathrm{log}_{10}\;\sigma_{e})/1.54,
\end{eqnarray}
respectively. The best-fit Coma relation slopes ($\alpha=1.30\pm0.08$, $\beta=-0.82\pm0.03$) and zero point ($\gamma=-0.443$) in the rest-frame B-band are adopted as our local reference orientation of the plane \citep[][see also Table \ref{tab:fit}]{Jorgensen+06}.

The best-fit Coma data, from \cite{Jorgensen+19}, was fit using the same method as described below while adopting the FP parameters from \cite{Jorgensen_Chiboucas2013} ($\alpha=1.30\pm0.08$, $\beta=-0.82\pm0.03$) to get the zero point  ($\gamma=-0.443$). We use this as our local reference cluster fit and list the parameters along with their uncertainties in Table \ref{tab:fit}.\\

In Figure \ref{fig:FP_eo_fo}, the FP edge-on and face-on projections, as described in Equation \ref{eq:FPeo} and \ref{eq:FPfo}, are shown. We use this to examine how $z\sim2$ massive quiescent galaxies populate and evolve to $z=0$ in this plane. In the edge-on FP, the dominating errors from the stellar velocity dispersion are shown on the y-axis. For the face-on plane, the errors are calculated similarly to the approximation used in \cite{Jorgensen+06}.

The COSMOS quiescent galaxies from S20 and B17 are found to be below (in the edge-on plane) and above (in the face-on plane) the local Coma FP relation. These galaxies have compact sizes and younger stellar populations (due to their high redshift and more recent quenching), effectively increasing their mean effective surface brightness. 

An edge-on FP cannot be clearly established using the S20 sample alone. However, when fitting the S20 and B17 samples together, a FP is in place at $1.5<z<2.5$ (hereafter referred to as $\rm{FP}_{z\sim2}$). The FP is fitted by using the least-squares method, minimizing the least-squares in the y-direction, with uncertainties from a bootstrapping method (see relation 2 in Table \ref{tab:fit}). In \cite{vandeSande2014_FP}, a FP was indicated for a similar epoch ($1.5<z<2.5$). However, in this study, the sample of $z>2$ galaxies is $3\times$ more numerous, robustly confirming the existence of a plane at $1.5<z<2.5$.

A $\rm{FP}_{z\sim2}$ is established using $19$ massive quiescent galaxies that as a result must have been a relatively homogeneous population already at this epoch signaling even earlier formation and significant evolution from $z=2$ to the present-day.


\section{Evolution of the Scaling Relations} \label{sec:SR_evol}

In Section \ref{sec:results} we established a FP for MQGs at $1.5<z<2.5$. Here, we explore how the $z=2$ MQGs from S20 evolve through these scaling relations to the CND-matched minimal progenitor biased (Section \ref{sec:MASSIVEnx}) local MASSIVE(n$_{17}$) sample. The $z=2$ MQGs was, in S20, shown to undergo structural ($\Delta{{\rm{log}}}r_{e,circ}\sim0.6$) and stellar mass ($\Delta{{\rm{log}}}M_{\ast}\sim0.3$) evolution with a stellar-to-dynamical mass ratio, $\Delta{}\rm{log}_{10}M_{\ast}/\Delta{}\rm{log}_{10}M_{\rm{dyn}} \sim 0.5$, from $z=2$ to $0$. These effects was in S20 suggested to arise from minor merger driven size growth. Additionally, the luminosity is expected to change due passive evolution of the stellar population and with the addition of new stellar mass from the minor merger driven size evolution. We adopt the structural and dynamical mass evolution from S20 and in Section \ref{sec:model_passive} and \ref{sec:lum_minor_mergers} model the before mentioned luminosity evolution components. Combining these effects we explore if they show a consistent picture of evolution when analysed in the scaling relations.\\

Finally, to highlight how the different physical mechanisms affect the evolution of the galaxies, throughout the scaling relations, Equation \ref{eq:logML_all} was derived. The evolution of the FP parameters can be formalised \citep{Saglia+10,Saglia+16_erratum2010} under the assumption of homology, where $\alpha,\beta$ are constant over time \citep{Beifiori+17},
\begin{equation} \label{eq:DlogL}
    \Delta{}\mathrm{log}_{10}\; L = \frac{1+2\beta}{\beta}\; \Delta{}\mathrm{log}_{10}\; r_{e} - \frac{\alpha}{\beta}\; \Delta{}\mathrm{log}_{10}\; \sigma_{e} - \frac{\Delta{}\gamma}{\beta}.
\end{equation}
The logarithmic difference is define by $\Delta{}\mathrm{log}_{10}X = \mathrm{log}_{10}( X_{z=2}/X_{z=0})$ where $X$, in this case, is either the luminosity, size or dispersion. The zero point evolution is described by $\Delta{}\gamma = \gamma_{z} - \gamma_{z=0}$.
This leads to the relation between the zero point evolution and the average change in ML, $\Delta{}\mathrm{log}_{10}\; M/L = \Delta{}\gamma/\beta$. We express the change in ML by effective size, stellar velocity dispersion, and luminosity evolution,
\begin{eqnarray} \label{eq:logML} 
   \Delta{}\mathrm{log}_{10}\; M/L &=& \frac{1+2\beta}{\beta}\; \Delta{}\mathrm{log}_{10}\; r_{e} - \frac{\alpha}{\beta}\; \Delta{}\mathrm{log}_{10}\; \sigma_{e} \nonumber \\ && - \Delta{}\mathrm{log}_{10}\; L_{\rm{total}}.
\end{eqnarray}
The change in effective size and stellar velocity dispersion are adopted from S20 as mentioned above. The total luminosity contribution from $z=2$ to $0$ can be described by both the luminosity change due to passive evolution, $\Delta{}\mathrm{log}_{10}\; L_{\rm{passive}}$, and the luminosity change associated with the newly incorporated stellar mass within the effective radius, $\Delta{}\mathrm{log}_{10}\; L_{M_{\ast}}$, 
\begin{eqnarray}
    \Delta{}\mathrm{log}_{10}\; L_{\rm{total}} = \Delta{}\mathrm{log}_{10}\; L_{\rm{passive}} + \Delta{}\mathrm{log}_{10}\; L_{M_{\ast}}.\\ \nonumber
\end{eqnarray}
The $\Delta{}\mathrm{log}_{10}\; L_{M_{\ast}}$ can be expressed in terms of the stellar mass and $M_{\ast}-L$ evolution,
\begin{eqnarray} \label{eqn:fdm}
    \Delta{}\mathrm{log}_{10}L_{M_{\ast}} &=& \Delta{}\mathrm{log}_{10}M_{\ast} - \Delta{}\mathrm{log}_{10}\frac{M_{\ast}}{L}\\
    &=& x\Delta{}\mathrm{log}_{10}M_{\rm{dyn}} - \Delta{}\mathrm{log}_{10}\bigg(\frac{M_{\ast}}{L}\bigg)_{M_{\ast}}.
\end{eqnarray}
Here $x=\Delta{}\rm{log}_{10}M_{\ast}/\Delta{}\rm{log}_{10}M_{\rm{dyn}}$ and $\Delta{}\mathrm{log}_{10}M_{\rm{dyn}}$ are adopted from S20 (see the first paragraph). The total change in ML can thus be expressed in terms of the passive (Section \ref{sec:model_passive}), structural (Section \ref{sec:model_struct}), stellar-to-dynamical mass ratio (Section \ref{sec:model_fdm}), and the luminosity contribution from the stellar mass increase due to minor merger driven size growth (Section 
\ref{sec:lum_minor_mergers}),
\begin{eqnarray} \label{eq:logML_all} 
   \Delta{}\mathrm{log}_{10}\; M/L &=& \frac{1+2\beta}{\beta}\; \Delta{}\mathrm{log}_{10}\; r_{e} - \frac{\alpha}{\beta}\; \Delta{}\mathrm{log}_{10}\; \sigma_{e} \nonumber \\ && - \Delta{}\mathrm{log}_{10}\; L_{\rm{passive}} \nonumber \\ && -  x\Delta{}\mathrm{log}_{10}M_{\rm{dyn}} + \Delta{}\mathrm{log}_{10}\bigg(\frac{M_{\ast}}{L}\bigg)_{M_{\ast}}.
\end{eqnarray}

We show the effect of the mean size and dispersion evolution from S20 and the modeled effect on the luminosity in Figure \ref{fig:FP_eo_model} and in Appendix \ref{app:figures1+2_wmodels}. The effects of passive (red) and structural (blue) evolution with the effect of the stellar-to-dynamical mass ratio (black) inclusion, and the luminosity increase from the stellar mass (green) is shown in arrows from the median of the S20 sample of $z=2$ MQGs. The effects of passive and structural evolution together with the change in dynamical-to-stellar mass ratio presents a realistic scenario of how these S20 MQGs could evolve into local galaxies.

\begin{figure}[t]
    \centering
        \includegraphics[width=8.5cm]{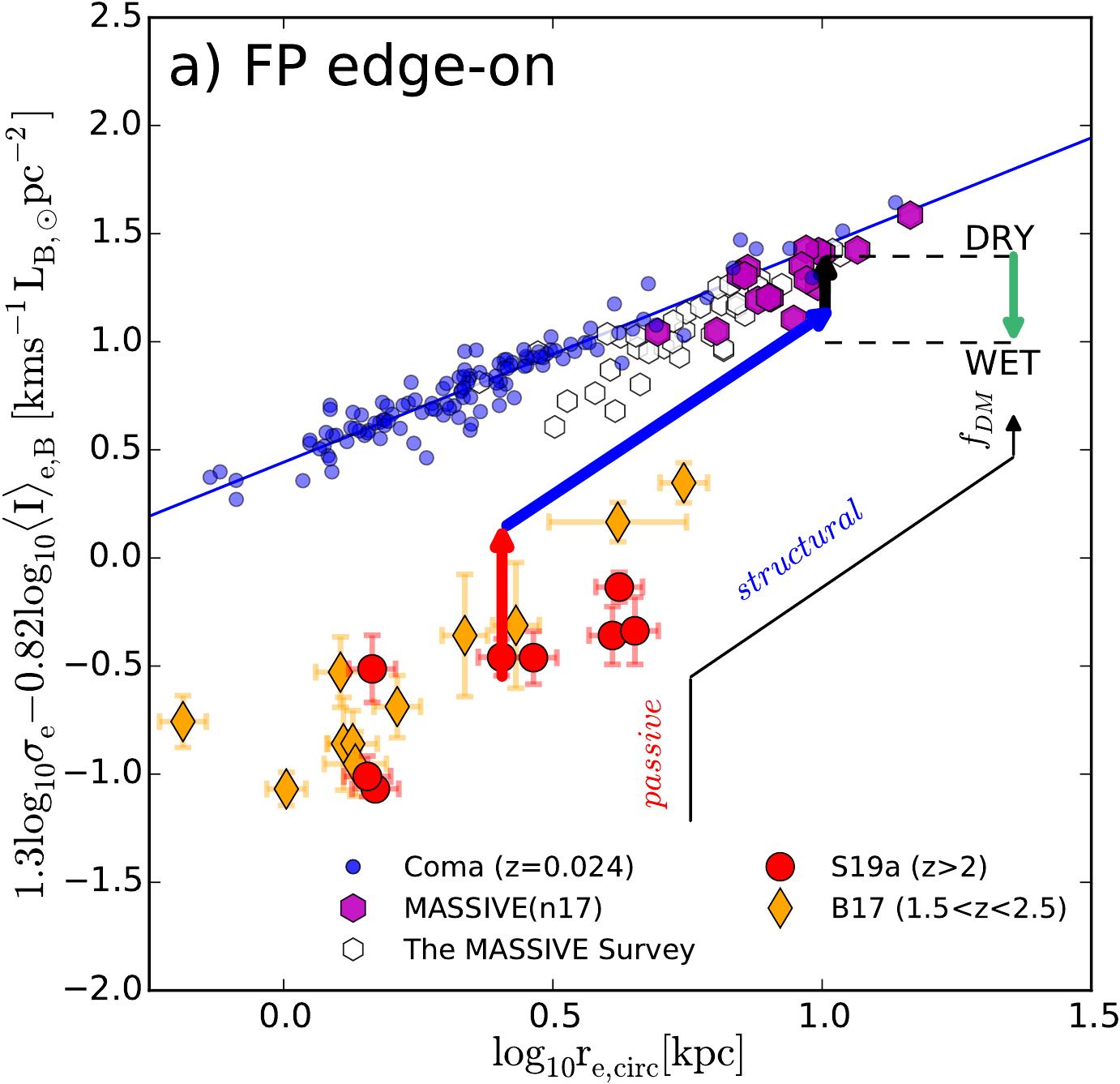}
        \caption{The FP projected edge-on (as portrait in Figure \ref{fig:FP_eo_fo}) shown with the predicted evolution (see Section \ref{sec:SR_evol}) of the S20 sample from $z=2$ to $0$. The average decrease in luminosity from passive evolution of the stellar population (Section \ref{sec:model_passive}) is shown by the red arrow. The size increase from structural evolution due to minor mergers are shown as the blue arrow (Section \ref{sec:model_struct}). The black arrow shows the effect of a changing $M_{\ast}/M_{\rm{dyn}}$ from $1$ to $0.5$ (Section \ref{sec:model_fdm}). The green arrow represents the surface brightness change for dry or wet minor merger-driven structural evolution (see Section \ref{sec:lum_minor_mergers}).
        The MQGs from S20 are consistent with evolving, via both passive and primarily dry minor merger structural evolution when taking the change in stellar-to-dynamical mass ratio into account, into the local most massive Coma galaxies and the CND-matched minimal progenitor biased MASSIVE(n$_{17}$) sample. These evolutionary trends are shown on Figures \ref{fig:ML_M_sigma} and \ref{fig:FP_eo_fo} in Appendix \ref{app:figures1+2_wmodels}.}
        \label{fig:FP_eo_model}
\end{figure}

\subsection{Passive Evolution} \label{sec:model_passive}
The expected passive evolution of the stellar population, from the redshift of formation to the present day, is estimated based on the evolution in $M_{\ast}/L$ of a BC03 \citep{BC+03} Simple Stellar Population (SSP) model with Chabrier IMF and solar metallicity. The formation redshift is estimated from the mass-weighted age and the redshift of observation. The passive evolution takes into account mass-loss during stellar evolution, and the $M_{\ast}$ is thus the mass locked into stars at a given age. The mass-weighted mean stellar age is indicative of the epoch by which the bulk of the stellar mass was formed. The $M_{\ast}/L$ uncertainties are dominated by the $1\sigma$ standard deviation of the age ($\sim0.2$ dex) when assuming solar metallicity. The definition of passive evolution assumes that no new stellar mass is added and thus $M_{\ast}/L$ directly relates to the $\Delta{}\mathrm{log}_{10}\; L_{\rm{passive}}$ from Equation \ref{eq:logML_all}.

The average estimate of the passive evolution is shown by the red arrow in Figure \ref{fig:FP_eo_model}. This is what is expected as Equation \ref{eq:logML_all} reduces to $\Delta{}\mathrm{log}_{10}\; M/L = \Delta{}\gamma/\beta = - \Delta{}\mathrm{log}_{10} L_{\rm{passive}}$ for 
no structural ($\Delta{}\mathrm{log}_{10} r_{e}=\Delta{}\mathrm{log}_{10} \sigma_{e}=0$) and no luminosity change (from a stellar mass increase). The passive evolution cannot fully explain the direct evolution, of MQGs from $z=2$ to $0$, to Coma galaxies at fixed dynamical mass (or size). Furthermore, it is inadequate in alone explaining the expected evolution to the MASSIVE(n$_{17}$) sample.\\
\mbox{}\\

\subsection{Structural evolution}
\label{sec:model_struct}
We learned from S20 that the size growth with shallow stellar velocity dispersion change was driven by minor merger structural evolution. Using this information we can simplify Equation \ref{eq:logML_all},
\begin{equation}
\Delta{}\mathrm{log}_{10} M/L = \frac{\Delta{}\gamma}{\beta} = \frac{1+2\beta}{\beta}\; \Delta{}\mathrm{log}_{10}r_{e}.
\end{equation}
The effect of the structural evolution adopted from S20 is based on the size increase from the stellar mass size plane, their figure 8 ($\Delta{{\rm{log}}}r_{e,circ}\sim0.6$) alongside insignificant stellar velocity dispersion evolution ($\Delta{{\rm{log}}}\sigma_{e}\sim0$). This effect is shown in the edge-on FP in Figure \ref{fig:FP_eo_model} as the sloped blue arrow. The combined effect of the mean passive evolution of the stellar population (red arrow) and mean the structural evolution from minor merger driven size increase (blue arrow) brings the S20 MQGs in the vicinity of the MASSIVE(n$_{17}$) sample.

\subsection{Stellar-to-dynamical mass ratio} \label{sec:model_fdm}
In S20 they found that the median stellar-to-dynamical mass ratio $\rm{x} = \Delta{}\rm{log}_{10}M_{\ast}/\Delta{}\rm{log}_{10}M_{\rm{dyn}} \sim 0.5$. Specifically the stellar mass doubles ($\Delta{{\rm{log}}}M_{\ast}\sim0.3$ dex) from $z=2$ to $0$. Following equation \ref{eq:logML_all}, we show how this ratio affects the evolution of the scaling relations when not unity,
\begin{equation}
    \Delta{}\mathrm{log}_{10}\; M/L= \frac{\Delta{}\gamma}{\beta} = \rm{x}\;\Delta{}\mathrm{log}_{10}M_{\rm{dyn}}
\end{equation}
Here the effect of a changing stellar-to-dynamical mass ratio can be seen to directly affect the change in the scaling relations. The effect for $x=0.5$ is shown in Figure \ref{fig:FP_eo_model}. When taking into account the stellar-to-dynamical mass ratio, that affect the conversion between the stellar and dynamical ML ratio we obtain a predicted evolution closer to the high mass end of the Coma relation and the median position of the MASSIVE(n$_{17}$) sample.

\subsection{Luminosity increase from wet minor merger stellar populations} \label{sec:lum_minor_mergers}
From S20 we learned that $z=2$ MQGs grow their size and stellar mass through minor mergers in their evolution to $z=0$. In addition to the decrease in the B-band luminosity, due to passive evolution, it is expected that such merger events could add to the B-band luminosity if these are star-forming galaxies at the time of merging. From now on, this type of merger is referred to as ``wet", contrary to the ``dry" minor mergers that are passive before merging. In Equation \ref{eq:logML_all} this reduces to the effect of the stellar ML change from new stars accreted from minor mergers,
\begin{eqnarray} \label{eq:logML_lumminormergers} 
   \Delta{}\mathrm{log}_{10}\; M/L &=& \Delta{}\mathrm{log}_{10}\bigg(\frac{M_{\ast}}{L}\bigg)_{M_{\ast}}.
\end{eqnarray}
The B-band luminosity increase from merging galaxies between $z=2-0$ is modeled with composite stellar population models from the BC03 library with solar metallicity to obtain the stellar ML (similar to Section \ref{sec:model_passive}). The star-formation history follows the evolution of the main-sequence \citep{Speagle+14}. It is assumed that, after merging, the galaxies stop forming stars and follow a passive evolution.

The median stellar mass increase ($\Delta{}{\rm{log}}M_{\ast}=0.3$ dex) from minor mergers, predicted in S20, are used assuming a 1:20 merger ratio. Note that in our simplistic model, the correct mass ratio does not play a significant role. We also investigate a more realistic scenario with mergers distributed across redshift (z=1.8-0.1) together with two extreme cases of all the mass added at $z=1.8$ or $0.1$. A B-band luminosity increase of $0.4-0.45$ dex is found in all cases (for more details see Appendix \ref{app:lum_model}).

In Figure \ref{fig:FP_eo_model} the effect of increasing luminosity, due to wet minor merger stellar populations, is shown by the green arrow ($0.4$ dex). This effect appears to be in disagreement with the location of the Coma relation and the MASSIVE(n$_{17}$) sample. This suggests that dry minor merger galaxies, with no additional luminosity increase, are a preferred evolution scenario for MQGs.


\section{Discussion} \label{se:discussion}

\subsection{Passive evolution of massive quiescent galaxies from $z=2$ to $0$} \label{sec:passive_evol_coma}

Studies of passive galaxies in $0.8<z<1.8$ clusters \citep[among others][]{Jorgensen+06,vanderMarel_vanDokkum+07,Jorgensen_Chiboucas2013,Beifiori+17,Jorgensen+19} find that the change in $M_{\ast}/L$ ratio can be explained by passive evolution to $z=0$. Below we explore if a similar analysis can account for the evolution of the scaling relations at $z\sim2$.

\cite{Jorgensen_Chiboucas2013} derived the $M_{dyn}/L_{\mathrm{B}}$ ratio evolution (based on models from \citealp{Maraston2005}), as a function of stellar age and metallicity, to be $\mathrm{log}M_{dyn}/L_{\mathrm{B}} = 0.935\ \mathrm{log\ age} + 0.337[\mathrm{M/H}] - 0.053$. Assuming passive evolution from $z=2$ to the best-fit Coma relation, at ${\rm{log}}(M_{\rm{dyn}}/M_{\odot})=11.5$, we find a formation redshift of $z_{\rm{form,Coma}}=2.01^{+0.1}_{-0.04}$ (for details see Appendix \ref{app:zform_ML}). The formation redshift is similar to the redshift of observation, which leaves too short a time to form the S20 MQGs at this epoch. The formation redshift, derived from our stellar population mass-weighted ages (assuming the median age), is $z_{\rm{form}}=3.41^{+4.92}_{-0.91}$. The uncertainties are estimated using the $1\sigma$ age uncertainties. Based on this, we conclude that the S20 MQGs at $z>2$ cannot evolve to the Coma relation by passive evolution alone.


\subsection{Minor merger-driven structural evolution of massive quiescent galaxies} \label{sec:evol_MASSIVE}
The fixed CND-matched MASSIVE(n$_{17}$) sample allows us to study the evolution of massive galaxy scaling relations from $z=2$ to $0$ with minimal progenitor bias. Evidence against purely passive evolution to $z=0$ is present in both the scaling relations (Figures \ref{fig:FP_eo_model} and \ref{fig:fig1+2_wmodels}) as explicitly shown in the previous section. The S20 MQGs at $z>2$ are consistent with evolving into the very most massive Coma galaxies, and the MASSIVE(n$_{17}$) sample, through passive, structural,  and stellar-to-dynamical mass ratio evolution.

The size increase of massive quiescent galaxies in cosmological simulations could be explained by adiabatic expansion due to AGN, decreasing the central mass density and puffing up the galaxies \citep{Dubois+13,Choi+18}. Major mergers, as the dominant mechanism for size growth, have become less popular as they make the galaxies too massive to be consistent with massive nearby galaxies. In S20, the structural evolution is interpreted to be from minor mergers in line with the scenario presented in the idealized simulations from \cite{Hopkins+09,Naab+09,Hilz+12,Hilz+13}. Here, the effective half-light radius grows by adding stars to the outskirts of the galaxy from tidally stripped minor mergers. This scenario is shown to cause inside-out growth, starting from a compact elliptical galaxy (core) that causes, through minor mergers, a build-up of the surface density profile wings, a present-day analog of a giant elliptical galaxy (core-envelope). A consequence of the inside-out minor merger growth scenario from \cite{Hilz+12} is an increasing dark matter fraction which has been suggested to cause a tilt in the FP over time \citep{Boylan-Kolchin+05,D'onofrio+13}. Essentially this is a consequence of using the effective radius as a foundation for analyzing the evolution, as this results in the stellar velocity dispersion mainly tracing the stars at $z=2$ but at $z=0$ a higher fraction of dark matter to stars. This effectively introduces a systematic difference between comparing stellar velocity dispersions across epochs that only trace stars with a similar sample where the stellar-to-dynamical mass ratio evolves. This systematic difference, included in Section \ref{sec:model_fdm}, must be taking into account and can be seen as the black arrow in Figures \ref{fig:FP_eo_model} and \ref{fig:fig1+2_wmodels}.


\subsection{Dry minor merger evolution}
In Section \ref{sec:lum_minor_mergers}, the predicted luminosity increase from wet minor mergers is modeled under the assumption that they are the primary drivers of the size growth. For the realistic scenario of adding wet minor mergers continuously from $z=2$ to $0$, we find that the luminosity increases by roughly $0.4$ dex.

In Figures \ref{fig:FP_eo_model} and \ref{fig:fig1+2_wmodels}, the predicted position of the S20 MQGs (following passive, structural, and stellar-to-dynamical mass evolution) is indicated alongside the effect of the luminosity from wet minor mergers. The green arrow indicates how the predictions would move compared to the local best-fit relation of Coma and around the locus of the MASSIVE(n$_{17}$) sample, strongly favoring dry minor mergers.

In the inside-out growth scenario, the rest-frame B-band luminosity increase takes place in the outer parts of the galaxy. The luminosity from the MASSIVE(n$_{17}$) sample is measured using SDSS deVMag which represents the luminosity of the galaxy out to $8r_{e}$. Thus, an underestimation of the luminosity, by only sampling the central part of the galaxy and missing the outskirts, is unlikely.

The wet minor merger luminosity increase offsets the predictions from the local relation and the MASSIVE(n$_{17}$) sample and thus appears to not be a favored way to grow MQG at $z\sim2$. Another possibility is that the minor mergers already have quenched stellar populations (before their merger) with low rest-frame B-band luminosity \citep{Oogi+13,Naab+14,Tapia+14}. The evolution from $z=2$ to $0$ of the FP and $M/L$ ratio scaling relations is consistent with such a scenario, caused primarily by dry minor mergers, passive, and stellar-to-dynamical mass ratio evolution for massive quiescent galaxies.



\subsection{Caveats} \label{sec:Caveats}

Data from S20 and B17 are combined to establish the $\rm{FP}_{z\sim2}$ at $1.5<z<2.5$ with more than half of the sources at $z>2$. A large fraction of the quiescent galaxies from B17 is found to be disk-like \citep[based on S\'ersic index, $n<2.5$][]{Belli+17}, which could mean that an unknown contribution from rotation is included in the measured stellar velocity dispersion. For spherical dispersion-dominated systems, the circularized radius and semi-major axis are comparable methods of size measurement. However, for more disk-like systems the difference grows between the two size measuring methods, further causing a bias between dispersion and rotation dominated galaxies. We estimate, based on the axis ratios, that the circularised sizes differ by $7-30\%$ compared to the semi-major axes. This is well within the quoted uncertainties of the predicted position of the S20 MQGs at $z=0$. This bias could potentially affect the zero point and coefficients of the best-fit in Figure \ref{fig:ML_M_sigma} and \ref{fig:FP_eo_fo}. This issue could be solved by spatially resolved spectroscopy disentangling the contribution from rotation and dispersion.\\

The SDSS modelMag luminosity of the most luminous galaxies have been underestimated \citep{Bernardi+17}. In our study, the deVMag and cmodelMag methods have been used to estimate the luminosity of the MASSIVE and Coma sample, respectively. The median offset for the brightest galaxies ($M_{r}\sim-24$), in the $r$-band, are $\sim15\ \%$ (see Figure 7 in \citet{Bernardi+17}). This translates to a difference in $\Delta{}\rm{log}_{10}L \propto \rm{log}_{10}(10^{0.4\Delta{}M_{r}}) = 0.4 \cdot 0.15 = 0.06$, assuming that the $r$-band magnitude are representative for the Bessel B-band used in our study. In the case that this assumption is valid, the ML and average effective surface brightness would change by $0.06$ dex, thus moving the local galaxies in the positive y-direction, by the same amount, in Figure 1, 2, and 3. Such an effect is minimal and would not affect the general trends, results, and conclusions made in this paper.

The dominating uncertainty of the mass-rank scatter from the CND-matching of the local MASSIVE(n$_{17}$) sample does not affect the conclusions of this study. Furthermore, if using a probabilistic CND-matching approach \citep[see][]{Wellons&Torrey+17}, this would increase the number of galaxies in the MASSIVE(n$_{17}$) sample from $17$ to $30$. In Figure \ref{fig:ML_M_sigma} and \ref{fig:FP_eo_fo} this would correspond to a greater number of white hexagons becoming purple, which causes no noticeable effects on the trends in the figures. On the other hand, if not all massive galaxies at $z>2$ have similar merger histories, descendants that become quiescent systems at late times would have been missed \citep[see e.g.][]{Naab+14}. A study of the average stellar population age within the effective radius of the most massive ($\sigma_{\ast} > 220$ km/s) MASSIVE Survey galaxies \citep{Greene+15} find their ages to be $>10$ Gyr. This suggest that our sample of MASSIVE(n$_{17}$) galaxies already were quenched at $z>2$, and likely have similar merger histories from $z=2$ to $0$. Studies also show that the most massive end of the stellar mass function (${\rm{log}}(M_{\ast}/M_{\odot})>11.5$) evolves very little, if at all between $0<z<2$ \citet[see e.g. figure 5 in][]{McLeod+20}.


\section{Summary and Conclusion} \label{sec:summary}

In this work, we present the highest redshift study of quiescent galaxy scaling relations with a sample size $2\times$ larger than previous studies at this redshift. The $M/L$ ratio of massive quiescent galaxies at $z>2$ is observed to be $\sim30-40\times$ smaller than the local Coma relation (at fixed dynamical mass) and it requires significant passive luminosity evolution to match the $z=0$ relation. In S20, the same galaxies are shown to undergo considerable structural evolution by quadrupling their sizes from $z=2$ to $0$, while their effective dispersion remains nearly unchanged. In this paper, the FP and $M/L$ ratio established scaling relations at $z\sim2$, and the expected structural and passive evolution, are explored for the S20 MQGs from $z=2$ to $0$. The main conclusions of this study are listed below:

\begin{itemize}
    \setlength\itemsep{2em}
    \item The FP and $M/L$ ratio relations are established at $z\sim2$. Compared to the local Coma cluster and the CND-matched MASSIVE(n$_{17}$) sample, the quiescent galaxies at high redshift are found to be both compact and rest-frame B-band brighter, the latter due to more recently quenched stellar populations. The position of the MASSIVE(n$_{17}$) sample broadly agrees with the best fit Coma relation for the most massive and largest galaxies (Figure \ref{fig:ML_M_sigma} and \ref{fig:FP_eo_fo}).

    \item Interpreting the $M/L$ ratio offset as purely passive evolution of the stellar population leads to a formation redshift of $z\sim2$, lower than the formation redshift inferred from the stellar population analysis of S20 MQGs, $z_{\rm{form}}=3.41^{+4.92}_{-0.91}$ (Section \ref{sec:passive_evol_coma}). As a result, the S20 MQGs are not consistent with their evolution into the local Coma FP and $M/L$ ratio scaling relations by passive evolution alone. 
    
    \item The S20 MQGs at $z\sim2$ are consistent with passive, structural, and stellar-to-dynamical mass ratio evolution into the most massive Coma galaxies and the minimal progenitor biased MASSIVE(n$_{17}$) sample (See Section \ref{sec:SR_evol} and Figures \ref{fig:FP_eo_model} and \ref{fig:fig1+2_wmodels}).
    
    \item In the case that the observed size evolution can be attributed entirely to minor mergers, the FP and $M/L$ ratio evolutions are consistent with the accretion of dry minor merger stellar populations. A scenario of wet minor mergers increases the rest-frame B-band luminosity by $0.4$ dex inconsistent with the evolution of MQG at $z\sim2$ into the local most massive Coma and the CND-matched MASSIVE(n$_{17}$) galaxies. 

\end{itemize}

M.S., S.T., C.G., and G.B. acknowledge support from the European Research Council (ERC) Consolidator Grant funding scheme (project ConTExt, grant number 648179). The Cosmic Dawn Center (DAWN) is funded by the Danish National Research Foundation under grant No. 140. Based on observations made with the NASA/ESA Hubble Space Telescope, obtained from the data archive at the Space Telescope Science Institute. STScI is operated by the Association of Universities for Research in Astronomy, Inc. under NASA contract NAS 5-26555. Support for this work was provided by NASA through grant number HST-GO-14721.002 from the Space Telescope Science Institute, which is operated by AURA, Inc., under NASA contract NAS 5-26555. M.S. thank Nina Voit for her encouragement and unparalleled love, help, and support. This research made use of Astropy (version 1.1.1),\footnote{http://www.astropy.org} a community-developed core Python package for Astronomy \citep{astropy:2013, astropy:2018}. I.J. is supported by the Gemini Observatory, which is operated by the Association of Universities for Research in Astronomy, Inc., on behalf of the international Gemini partnership of Argentina, Brazil, Canada, Chile, the Republic of Korea, and the United States of America. M. H. acknowledges financial support from the Carlsberg Foundation via a Semper Ardens grant (CF15-0384). FV acknowledges support from the Carlsberg Foundation research grant CF18-0388 ``Galaxies: Rise And Death''.


\appendix

\section{The derivation of the luminosity and effective surface brightness, $\langle{} I\rangle{}_{\MakeLowercase{e},\mathrm{B}}$}
\label{app:Ie}

The luminosity and average effective surface brightness are estimated by converting the EAZY \citep{Brammer+08} calculated rest-frame B-band fluxes to apparent AB magnitudes (assuming no extinction correction) and calculating the absolute Vega magnitudes and luminosity by
\begin{equation} \label{eqn:Ie_Mabs}
M_{Vega,\mathrm{B}} = m_{Vega,\mathrm{B}} - 5\cdot({\rm{log}}(D_L/{\rm{pc}})-1)\;\;\;\; \frac{L_{\mathrm{B},{\rm{gal}}}}{L_{\mathrm{B},\odot}} = 10^{-0.4({M_{Vega,\mathrm{B}}-M_{\odot,\mathrm{B}})}}.
\end{equation}
Here the luminosity distance ($D_L$) and $M_{\odot,\mathrm{B}} = 5.45$\footnote{\url{http://mips.as.arizona.edu/~cnaw/sun.html}} are used. The effective surface brightness in Bessel B-band is calculated as
\begin{equation} \label{eqnapp:Ie_eqn}
\langle I\rangle_{e,\mathrm{B}} = \frac{L_{\mathrm{B},{\rm{gal}}}/L_{\mathrm{B},\odot{}}}{2\pi{}r_{e}^2}.
\end{equation}
Note that cosmological redshift dimming is included when converting the radius from arcsec to parsec.

\section{How representative is the MASSIVE(n$_{17}$) sample} \label{app:massiven}

\begin{figure*}[t]
    \centering
        \includegraphics[width=18cm]{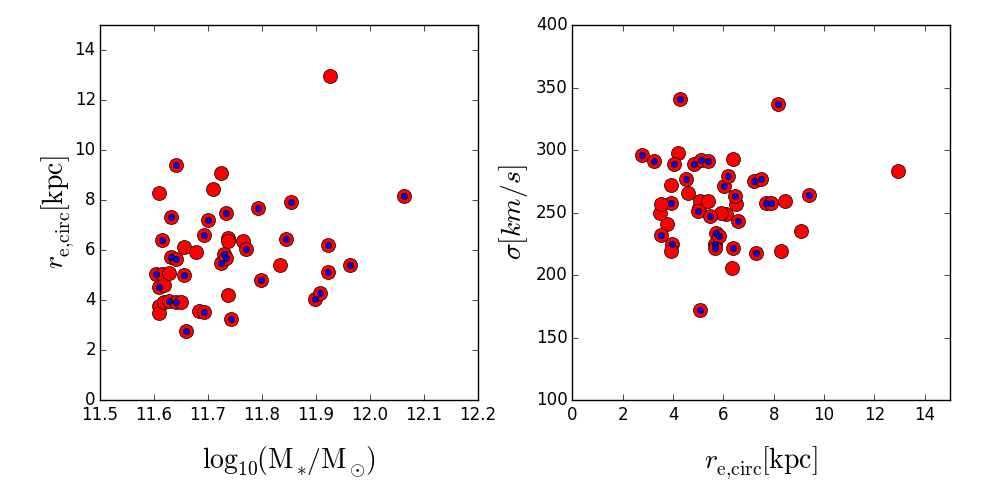}
        \caption{The distribution of MASSIVE(n) (red) and MASSIVE(n$_{17}$) (blue) for the stellar mass-size and size-dispersion plane. The MASSIVE(n$_{17}$) sample, with available SDSS photometry, is selected uniformly from the parent MASSIVE(n) sample and can be considered representative for the CND-matched parent sample.}
        \label{fig:MASSIVEnx}
\end{figure*}

To ensure that we do not introduce a bias we compare the MASSIVE(n$_{17}$) and the parent MASSIVE(n) sample's stellar mass, size or stellar velocity dispersion. This is presented in Figure \ref{fig:MASSIVEnx} where we show that the MASSIVE(n$_{17}$) sample is representative of the structural and kinematical parameters.

\section{The evolution of the FP and ML relations} \label{app:figures1+2_wmodels}
\begin{figure*}[p]
    \centering
        \includegraphics[width=18cm]{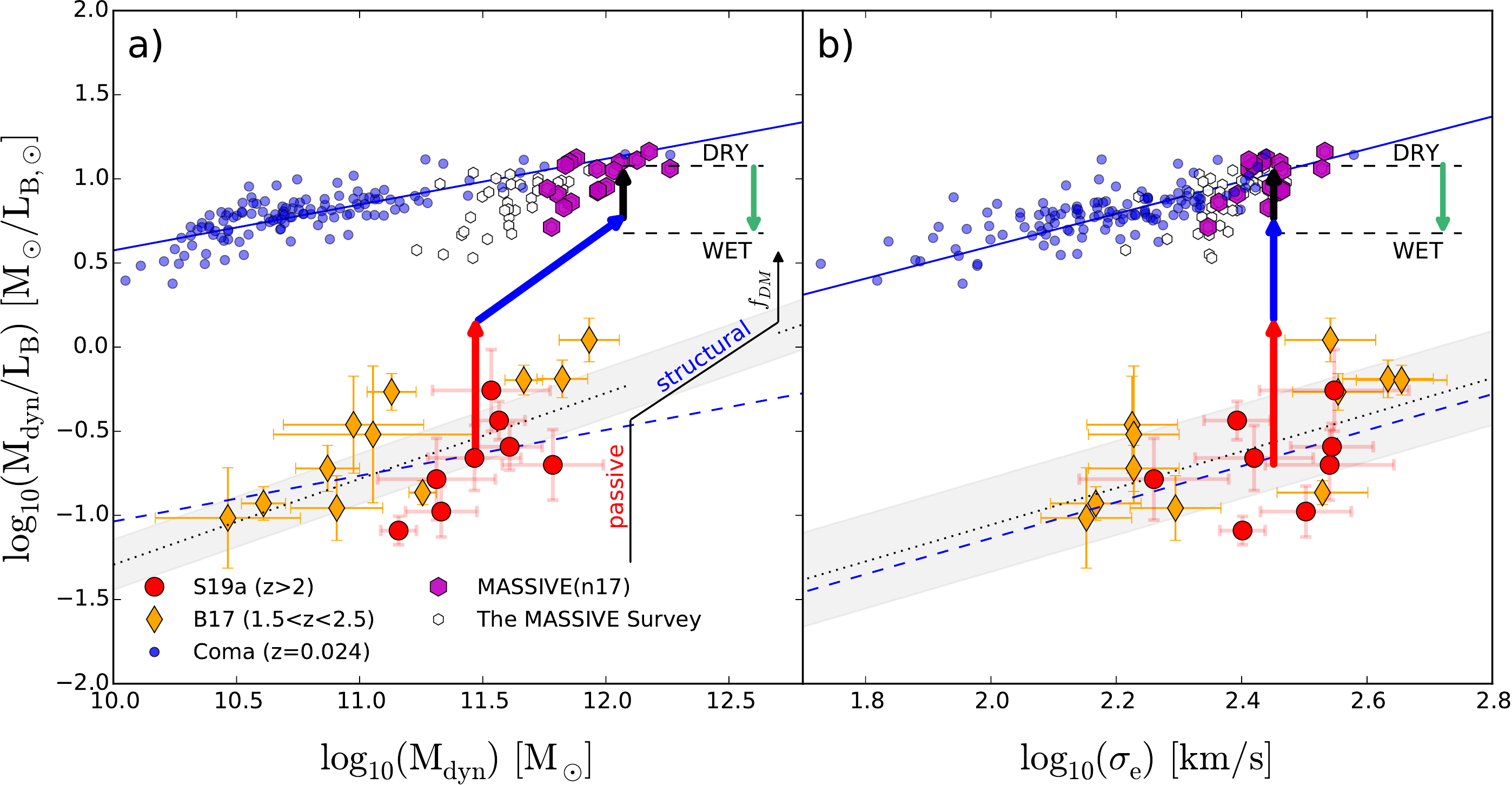}
        
        \hfill \mbox{}\\
        \hfill
        
        \includegraphics[width=.5\textwidth]{FP_eo_nosim_CNDfix_wmodel_devrad_v1.pdf}
        \includegraphics[width=.48\textwidth]{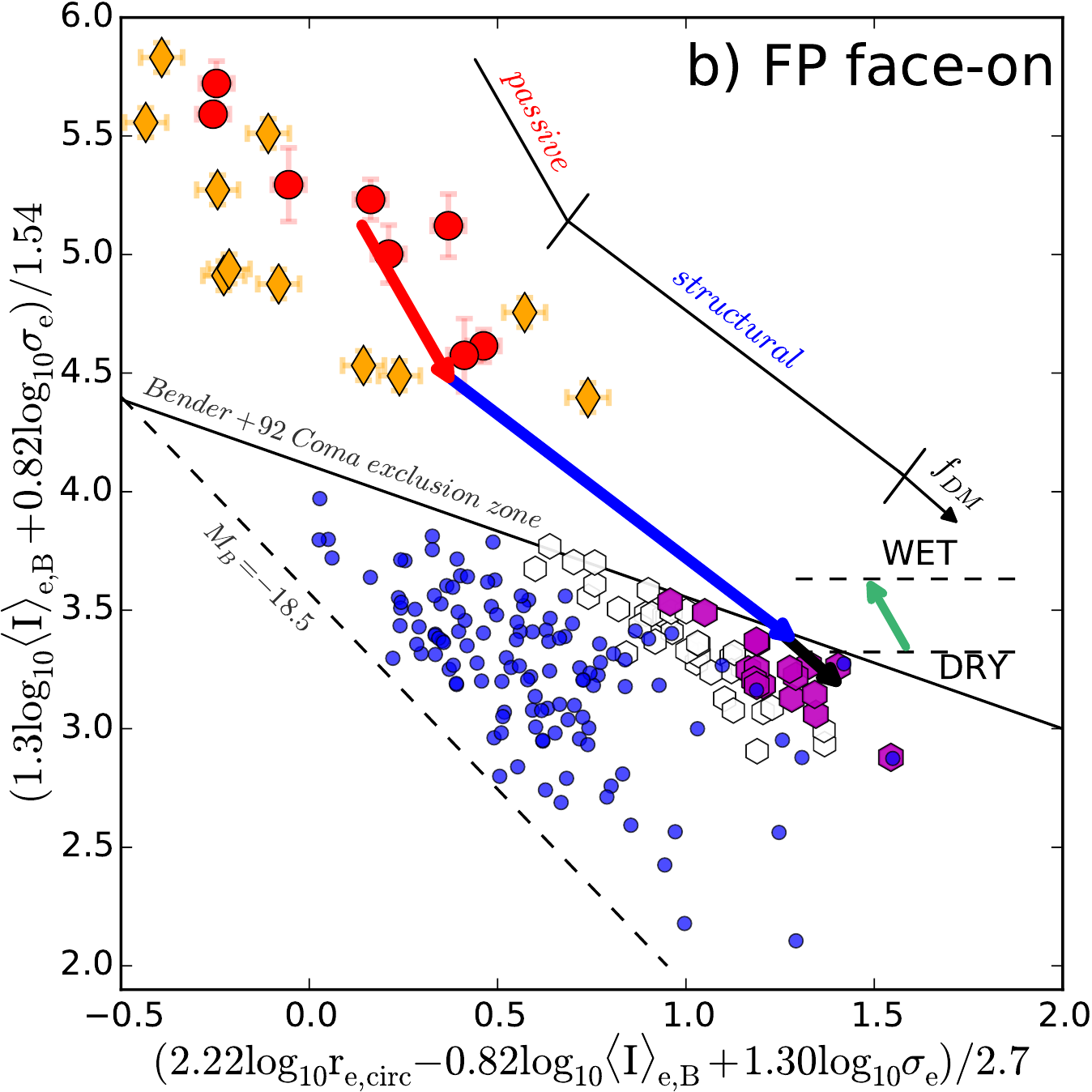} 
        \caption{The FP and ML scaling relations from Section \ref{sec:results} are shown with the predicted passive, structural, dark matter fraction evolution (red, blue and black arrow) and luminosity change due to minor mergers (green arrow). \label{fig:fig1+2_wmodels}}
\end{figure*}

We present Figure \ref{fig:fig1+2_wmodels} showing Figures \ref{fig:ML_M_sigma} and \ref{fig:FP_eo_fo}, from Section \ref{sec:results}, here with the mean evolutionary trends covered in Section \ref{sec:SR_evol}.

\section{Details on the modeling of the B-band luminosity increase due to minor merger added stellar populations} \label{app:lum_model}

\begin{figure}[t!]
    \centering
        \includegraphics[width=8.5cm]{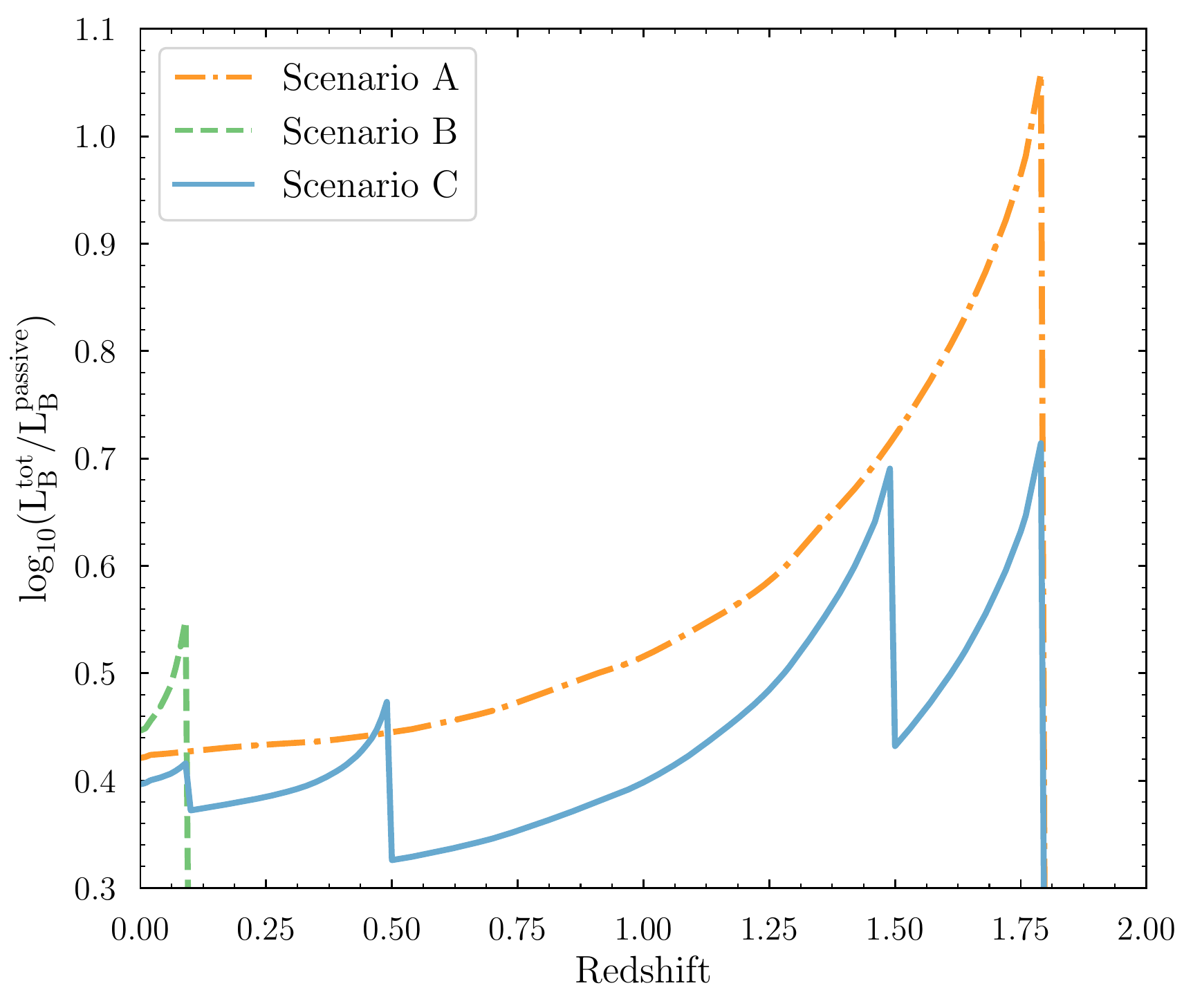}
        \caption{Change in B-band luminosity of different galaxies during passive evolution (after turn-off of star-formation) after evolving along the star-forming main-sequence. The luminosity is compared to a galaxy on the main-sequence until $z = 0$. The minor mergers added are match to the predicted stellar mass increase. Model C, the most realistic one adds relative fractions $0.1 / 0.2 / 0.3 / 0.4$ of minor mergers at a redshift of $z=0.1 / 0.5 / 1.5 / 1.8$, where as model A and B assumes the extremes of adding all the stellar populations at either $1.8$ and $z=0.1$, respectively. The luminosity increase is scaling directly with the added stellar mass and is independent of the mass ratio. The final relative luminosity increase due to minor merger-driven structural evolution is not strongly affected by adding stellar mass at different epochs and give values of $\rm{log}_{10}(L_{B}^{tot}/L_{B}^{passive})\sim0.4$.}
        \label{fig:lum_increase_wet}
\end{figure}

The amount of B-band luminosity increase due to the minor wet and dry newly added merging stellar populations are constrained between redshifts $z=2$ to $0$, based on simple assumptions. Figure \ref{fig:lum_increase_wet} shows the B-band luminosity increase due to minor mergers (on top of the luminosity decrease due to passive evolution) as a function of redshift for three scenarios.
In scenario A, it is assumed that all the merging happens about $300$ Myrs after the galaxies are observed at $z=1.8$. In scenario B, the galaxies merge at z=0.1. Note that, since the merging galaxies follow the global star-forming main-sequence, and hence have lower SFRs at lower redshifts on average, the increase in luminosity is less at $z=0.1$ than at $z=1.8$. Finally, scenario C shows a more realistic merger history for which $10\ \%, 20\ \%, 30\ \%$, and $40\ \%$ of the stellar mass increase happens via mergers at $z=0.1, 0.5, 1.5$, and $1.8$, respectively. These follow roughly the measured trends of merger fraction in the literature \citep[e.g.][]{Man+12,Newman+12,Man+16}. Although the merger history in the different scenarios is very different, the final increase in rest-frame B-band luminosity is very similar between $0.4$ and $0.45$ dex.

\section{Formation redshift from M/L relation} \label{app:zform_ML}
\cite{Jorgensen_Chiboucas2013}, using \cite{Maraston2005} models, predict the $M/L_{\mathrm{B}}$ ratio evolution as a function of age and metallicity \citep[Table 9]{Jorgensen_Chiboucas2013}
\begin{equation}
\mathrm{log}M/L_{\mathrm{B}} = 0.935\ \mathrm{log\ age} + 0.337[\mathrm{M/H}] - 0.053.
\end{equation}
For passive evolution with constant metallicity, the difference in ${\rm{log}}_{10}M/L_{\rm{B}}$ can be related to the age of the stellar population.
\begin{equation}
\Delta{}\mathrm{log}M/L_{\mathrm{B}} = 0.935\ \Delta{}\mathrm{log\ age} \label{eq:DlogML_age}
\end{equation}
If the MQGs at $z>2$ are the progenitors of the local Coma relation, the change in $M/L_{\rm{B}}$ (at fixed dynamical mass) can be used to estimate a corresponding formation time. The age difference can be written in terms of look-back times and expressed as the formation time
\begin{eqnarray}
\Delta{}\mathrm{log\ age} &=& \mathrm{log\ age}_{\mathrm{z=0}} - \mathrm{log\ age}_{z\sim2} = \mathrm{log}(t_{\mathrm{form}}-t_{\mathrm{obs},z=0}) - \mathrm{log}(t_{\mathrm{form}}-t_{\mathrm{obs},z\sim2}) = \mathrm{log} \bigg( \frac{1-t_{\mathrm{form}}/t_{\mathrm{obs},z=0}}{1-t_{\mathrm{form}}/t_{\mathrm{obs},z\sim2}} \bigg)\\
\Leftrightarrow{} t_{\mathrm{form}} &=& \frac{10^{\Delta{}\mathrm{log\ age}}\cdot{}t_{\mathrm{obs},z\sim2} - t_{\mathrm{obs},z=0}}{10^{\Delta{}\mathrm{log\ age}}-1} = \frac{10^{(\Delta{}\mathrm{log}M/L_{\mathrm{B}})/0.935}\cdot{}t_{\mathrm{obs},z\sim2} - t_{\mathrm{obs},z=0}}{10^{(\Delta{}\mathrm{log}M/L_{\mathrm{B}})/0.935}-1}
\end{eqnarray}
The uncertainty on the formation redshift is estimated by varying the $M/L$ ratio uncertainties ($\sim0.25$ dex).\\
\mbox{}\\ \mbox{}\\ \mbox{}\\  \mbox{}\\ \mbox{}\\   \mbox{}\\ \mbox{}\\


\bibliographystyle{aasjournal}{}
\bibliography{Citations}{}


\end{document}